\documentclass[secnumarabic,amssymb, nobibnotes, aps, prd, 11pt, abstract]{revtex4-2}
\usepackage{tocloft}
\usepackage[dvips]{graphicx}
\usepackage{amsmath}
\usepackage{subcaption}
\usepackage{hyperref}
\usepackage{filecontents}
\hypersetup{colorlinks,linkcolor={red},citecolor={blue},urlcolor={red}}
\usepackage{setspace}
\usepackage{xcolor}
\pagecolor{white}
\usepackage{float}
\graphicspath{{./plots/}}

\usepackage{multirow}
\color{black}

\begin{document}

\singlespacing

	\title{Entanglement Islands, Page curves and Phase Transitions of Kerr-AdS Black Holes}%

	\author{Digen Das$^{1,2}$}
	
	\email{$rs_digendas@dibru.ac.in$}
	
	\author{Mozib Bin Awal$^1$}
	
	\email{$rs_mozibbinawal@dibru.ac.in$}

	\author{Prabwal Phukon$^{1,3}$}
	\email{$prabwal@dibru.ac.in$}
	
	\affiliation{$^1$Department of Physics, Dibrugarh University, Dibrugarh, Assam,786004.\\$^2$ Department of Physics, Sadiya College, Chapakhowa, Assam,786157.\\$^3$Theoretical Physics Division, Centre for Atmospheric Studies, Dibrugarh University, Dibrugarh,Assam,786004.\\}

	\begin{abstract}
	
We study the Page curve and information paradox for Kerr AdS black hole in light of entanglement entropy by employing the recently proposed island paradigm. By incorporating the island rule, we show that the entanglement entropy of Kerr AdS black hole grows linearly at early times and  declines to a constant value at late times in agreement with the well established Page curve. The novelty of this study resides in the investigation of influence of phase transitions on the page curve in two different ensembles. We find that a first order phase transition results in a sharp discontinuity in the Page curve. We study the evaporation process in different scenarios and find that in all the situations, the Page curve doesn't violate the  unitary principle of quantum mechanics.

\end{abstract}
	
	\maketitle
	
\section{Introduction}\label{sec1}

The ground breaking work of Hawking established that  a black  hole can emit particles with a thermal spectrum of temperature through a process named as Hawking radiation \cite{hawking1975particle, wald1975particle}. According to this, a black hole formed from collapse of pure matter state essentially evaporates into Hawking Radiation which is a maximally mixed state \cite{unruh1976notes}. However, this process violates the unitary principle of quantum mechanics and  it was observed that information may lost during the evaporation process  \cite{hawking1976breakdown}. This idea led to one of the most fundamental problem in theoretical physics known as the information paradox. It is believed that a clear understanding of the information paradox can be useful towards a theory of quantum gravity.
\par 
The information paradox can be addressed in light of entanglement entropy. As per calculation of Hawking \cite{hawking1976breakdown}, during the evaporation process, the entropy of Hawking Radiation increases monotonically and  it exceeds the thermodynamic entropy of the black hole at late time. If the black hole is initially prepared in a pure state, the principle of unitary evolution in quantum mechanics implies that the entanglement entropy of the black hole must be equal to that of the emitted radiation \cite{wang2024entanglement}. If we consider the entropy calculated by Hawking as the entanglement entropy of radiation, then it suggests that the entanglement entropy of the black hole exceeds the thermodynamic entropy of the black hole \cite{almheiri2021entropy}. This contradicts the principle that the entanglement entropy of the black hole should not exceed the thermodynamic entropy. However, if the evaporation process follows the unitarity of quantum mechanics, it should decline to zero at late time as the black hole vanishes.
\par
In \cite{page1993information,page2013time}, Don Page suggested that the evaporation of the black hole can be modeled as a state of random pure state and obtained the so called Page curve. He showed that  Thus the information paradox can  be connected to the evolution of the entanglement entropy of the Hawking radiation. Successful reproduction of the Page curve from the time evaluation of the entanglement entropy suggests a possible resolution to the information paradox.  The discovery of AdS/CFT correspondence \cite{maldacena1999large} provides a strong evidence for conservation of information in the evaporation process of black holes. It states that  quantum theory of gravity in Anti-de Sitter (AdS) space is dual to a conformal field theory at its boundary. Thus, a unitary process of the dual conformal field theory on the boundary can describe the black hole evaporation in AdS space. 
\par  Recently, significant progress has been made in computing the entanglement entropy of Hawking radiation and reproducing the Page curve within the framework of the AdS/CFT correspondence \cite{ almheiri2019entropy,  penington2022replica, almheiri2020replica, penington2020entanglement,almheiri2020page}. This development builds primarily on the pioneering work of Ryu and Takayanagi (RT) \cite{ryu2006holographic}, which relates the entanglement entropy of a boundary region to the area of a minimal surface in the bulk spacetime. Upon incorporating quantum corrections, the RT prescription was further generalized to the concept of quantum extremal surfaces \cite{barrella2013holographic, hubeny2007covariant, faulkner2013quantum,engelhardt2015quantum,lewkowycz2013generalized}. This generalized framework provides a systematic method for calculating the entanglement entropy of Hawking radiation.  This is commonly known as the island rule according to which the entanglement entropy of Hawking radiation is given by the following formula:
\begin{equation}
\begin{split}
     S_\text{Rad} &= \text{min}(\text{ext}[S_\text{gen}]) \\
    S_\text{gen} &= \frac{Area (\partial I)}{4 G} + S_\text{field} (R \cup I) \label{eq1}
\end{split}
\end{equation}
where, R is the radiation, I is the island (region inside the black hole), $\partial I$ is the boundary of the island, $S_{\text{field}}$ denotes the entanglement entropy of the fields in both the radiation and the island region. In the above equation (\ref{eq1}) $S_{\text{gen}}$ is the generalized entropy. The location of the island can be found by extremising the generalised entropy with respect to position and time\cite{carlip2014black}. One has to take the minimum value if there exist more than one extremum. Even without invoking holography, the island formula can be derived by means euclidean path integral on the replicated manifold wormholes \cite{penington2022replica, almheiri2020replica}. Moreover, this derivation of island formula is also valid for evaporating black holes \cite{wang2024entanglement, goto2021replica, hartman2020islands}.   Due to fact that the calculation of entanglement entropy of matter fields in four  or higher dimensional system  is highly non trivial, we employ the s-wave approximation \cite{hashimoto2020islands}. It is well established that, in the near-horizon region of a Kerr–AdS black hole, scalar field dynamics can be reduced to an effective two-dimensional theory. Following the dimensional reduction procedure as in \cite{jiang2007hawking}, we reduce the scalar field theory in four-dimensional Kerr–AdS spacetime to an effective 2D field theory. 
\par Though the island rule was originally motivated by investigations of two-dimensional black holes in JT  gravity, it has been widely employed to study a broad class of black hole spacetime, various asymptotic structures, and even extensions to higher-derivative theories of gravity \cite{hashimoto2020islands, almheiri2020entanglement, gautason2020page, anegawa2020notes, kim2021entanglement, karananas2021islands, hollowood2020islands, alishahiha2021island, geng2020massive, li2020reflected, goto2021replica, chandrasekaran2020including, bak2021unitarity, krishnan2021critical, chen2021evaporating, sybesma2021pure, ling2021island, matsuo2021islands,  wang2021islands,  wang2021page, li2021island, chu2021page,lu2022islands, yu2022islands, ahn2022islands, nam2022entanglement, afrasiar2023islands, anand2023page,  goswami2022small, yu2023entanglement, guo2023page, wu2023islands, wang2023entanglement, chowdhury2023mutual, PhysRevD.109.104053, Yu:2023whl, Chou:2023adi, Anand:2023ozw, Bousso:2023kdj, Jeong:2023lkc, Yadav:2022jib, Yadav:2023sdg, Bhattacharya:2021dnd}. However, there are few reports on rotating black hole. In these studies, nearly all of them studied the asymptotically flat black hole system. The island rule and Page curve for an asymptotically AdS black hole is studied in \cite{Lin:2024gip}. We address the information paradox problem in Kerr AdS black hole and this work may provide further insights in understanding black holes in  asymptotically AdS space time.
\par We study the influence of first order phase transition on the Page curve in eternal Kerr AdS black hole by considering canonical and an alternative new ensemble. In addition, We investigate the impact of the rotation parameter on the Page curve behavior for the same. We also modeled the evaporation of Kerr AdS black hole by introducing coupled baths. In particular, we analyzed how varying the rotation parameter changes the time evolution of the entanglement entropy for Large, Intermediate and Small black hole in both the ensembles.
\par The structure of this paper is arranged as follows: In Sec (\ref{sec2}) we provide brief review of Kerr AdS black hole and its dimensional reduction to an effective 2D theory. In Sec (\ref{sec3}) we calculate the entanglement entropy in both the situations viz. with and without island. Additionally, we calculate the Page time and Scrambling time and plot the Page curve. In Sec (\ref{sec4}) we demonstrate the effects of phase transition on the Page curve curve in two ensembles and finally in Sec (\ref{sec5}) we conclude our findings.
\section{Review of Kerr AdS black hole and its dimensional reduction}\label{sec2}
The metric of Kerr AdS black is given by \cite{carter1968hamilton, plebanski1976rotating}-
\begin{equation}
ds^2
= -\frac{\Delta_r}{\rho^2}
\left(
dt - \frac{a\sin^2\theta}{\Xi} d\phi
\right)^2
+ \frac{\rho^2}{\Delta_r} dr^2
+ \frac{\rho^2}{\Delta_\theta} d\theta^2
+ \frac{\Delta_\theta \sin^2\theta}{\rho^2}
\left(
a\, dt - \frac{r^2 + a^2}{\Xi} d\phi
\right)^2 ,
 \label{eq2}
\end{equation}

where
\begin{align*}
\Delta_r &= (r^2 + a^2)(1 + r^2 l^{-2}) - 2 M r = (r-r_{+})(r-r_{-}), \\
\Delta_\theta &= 1 - a^2 l^{-2} \cos^2\theta, \\
\rho^2 &= r^2 + a^2 \cos^2\theta, \\
\Xi &= 1 - a^2 l^{-2}.
\end{align*}

By setting $\Delta_r(r_H) = 0$ we can determine the horizon.
Following \cite{jiang2007hawking}, we can use a dimensional reduction technique to show that  the scalar field theory in a four-dimensional Kerr--AdS black hole can be reduced to that
of an effective two-dimensional spacetime near the horizon.
The action of a massless scalar field in the background spacetime
(\ref{eq2}) is
\begin{align}
S[\varphi]
&= \frac{1}{2} \int d^4x \sqrt{-g}\,
\varphi \nabla^2 \varphi \nonumber \\
&= \frac{1}{2}
\int dt\,dr\,d\theta\,d\phi\,
\frac{\sin\theta}{\Xi}\varphi
\Bigg[\frac{\Delta_{r} a^2 \sin^2\theta -\Delta_\theta (r^2 + a^2)^2}{\Delta_\theta \Delta_{r}} \partial_t^2
+  \frac{\Delta_r - \Delta_\theta (r^2 + a^2)}{\Delta_r \Delta_\theta} 2 a \Xi \partial_t \partial_\phi \nonumber \\
&\qquad
 + \frac{\Delta_r - \Delta_\theta a^2 \sin^2\theta}{\Delta_r \Delta_\theta \sin^2 \theta} \Xi^2 
\partial_\phi^2
+ \partial_r(\Delta_r \partial_r)+ \frac{1}{\sin\theta} \partial_\theta(\sin\theta\,\Delta_\theta \partial_\theta)
\Bigg]\varphi .
 \label{eq3}
\end{align}

Performing the partial wave decomposition of the scalar field
$\varphi = \sum_{l,m} \varphi_{lm}(t,r)\,Y_{lm}(\theta,\phi)$,
at the horizon, and transforming to the tortoise coordinate,
defined by
\begin{equation}
dr_* = \frac{r^2 + a^2}{\Delta_r} dr \equiv f(r)^{-1} dr \label{eq4}
\end{equation}
\begin{equation}
    r_{*} (r) = r + \frac{(r^2_{+}+a^2) \log(r-r_{+}) + (r^2_{-}+a^2) \log(r-r_{-}) }{r_{+}-r_{-}}
\end{equation}
The action (\ref{eq3}) can be simplified as
\begin{align}
S[\varphi]
&= \frac{r^2 + a^2}{2 \Xi}
\sum_{l,m} \int dt\,dr\, \varphi^*_{lm}
\Bigg[-
\frac{1}{f(r)}
\left(
\partial_t + i\frac{\Xi a m}{r^2 + a^2}
\right)^2
+ \partial_{r}(f(r)\partial_{r})
\Bigg]\varphi_{lm}.
 \label{eq5}
\end{align}
Therefore, by applying a near-horizon dimensional reduction procedure, each partial wave mode of the scalar field $\phi$ in a four-dimensional Kerr-AdS black hole can be equivalently represented as an infinite set of complex scalar fields propagating in an effective $(1+1)$-dimensional spacetime, coupled to a dilaton field $\Psi$ and a $U(1)$ gauge field $A_\mu$.

\begin{equation}
g_{tt} = -f(r) = -\frac{\Delta_r}{r^2 + a^2},
\qquad
g_{rr} = \frac{1}{f(r)}, \quad \psi = \frac{r^2 + a^2}{\Xi},
\qquad
A_t = - \frac{\Xi a}{r^2 + a^2},
\qquad
A_r = 0 .
\end{equation}
\par Thus following the above set of procedure, we can write an effective 2D metric for 4D kerr Ads Space time as - 

\begin{align}
     ds^2_{eff} = -f(r) dt^2 + f(r)^{-1} dr^2 = - \frac{(r-r_{+})(r-r_{-})}{r^2 + a^2 } dt^2 + \frac{r^2 + a^2}{(r-r_{+})(r-r_{-})}  dr^2
\end{align}

\par In the following chapters,we will proceed with the effective 2D geometry to investigate the information paradox for kerr AdS black hole. Thus the entanglement entropy of Hawking radiation in a 4D space time can be calculated by using the effective 2D space time framework. In case of Kerr AdS both superradiance and Hawking radiation contributes to the emitted radiation. The information paradox for a rotating BTZ black hole by considering the superradiance can be found in \cite{yu2022island} . However, in this study we will focus on the scenario where Hawking radiation dominates over superradiance. Thus in this case, we will work in small angular momentum limit i.e. $a<<M$.

\par In asymptotically AdS spacetime, Hawking radiation emitted by an AdS black hole get reflected back toward the horizon because of existence of an infinite potential in the boundary.  Thus, AdS effectively acts as a confining potential well. For sufficiently large AdS black holes, the reflected radiation returns to the black hole on a timescale short compared to the evaporation time. This leads to thermal equilibrium and a stable black hole with nonzero temperature. To allow for evaporation, one need to couple the system with a bath to collect the radiation.  We follow the construction given in \cite{almheiri2019entropy} and couple the AdS black hole to a non gravitational bath at finite temperature. In this setup, the black hole equilibrates with a bath located at the AdS boundary and subsequently undergoes  evaporation.
\begin{figure}[t]
    \centering
        \includegraphics[width=0.45\textwidth]{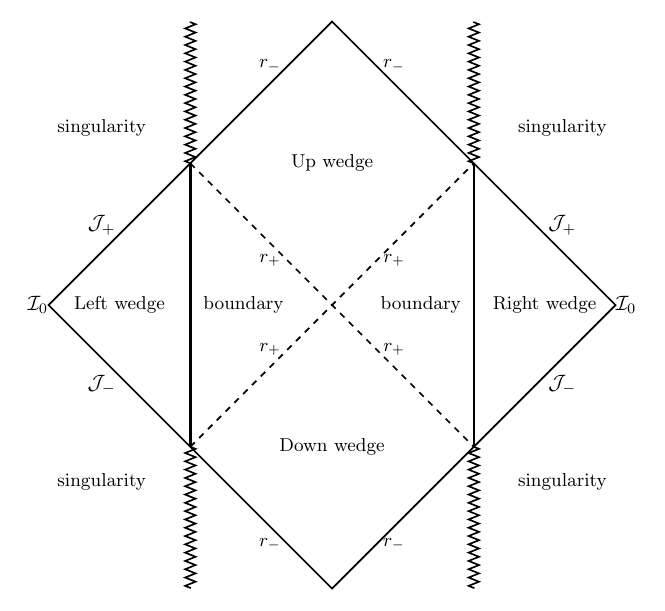}
        \caption{The Penrose diagram of a non extremal Kerr $\text{AdS}_4$ black hole. Here, $r_{\pm}$ denotes the outer and inner horizon respectively, $\mathcal{J}_{\pm}$ is future and past null infinity respectively, and $\mathcal{I}_{o}$ is spacelike infinity}
        

    \label{Penrose_genral}
\end{figure}
\subsection{TORTOISE COORDINATE FOR KERR ADS BLACK HOLE}

We employ the Kruskal co-ordinate transformation in order to perform the subsequent calculations of the entanglement entropy. The tortoise co-ordinate is defined as equation (\ref{eq4}). We then define the null co-ordinates $u= \tau - r_{*}$, $v= \tau + r_{*} $ and accordingly the Kruskal coordinates are written as - 
\begin{align}
    \text{Left Wedge}: U \equiv -e^{-\kappa u}, \qquad V \equiv e^{\kappa v} \\
    \text{Right Wedge}: U \equiv -e^{-\kappa u}, \qquad V \equiv e^{\kappa v}
\end{align}
In this way, we can eliminate the coordinate singularities. The penrose diagram for Kerr AdS black hole in Kruskal Co-ordinates is shown in Fig. (\ref{Penrose_genral}). Now under this transformation, the reduced metric can be written as a conformal flat metric as - 
\begin{equation}
    ds^2 = - \Omega^2 (r) dU dV 
\end{equation}
where, $\Omega^2 (r)$ is the conformal factor and is given by 
\begin{equation}
    \Omega (r) \equiv \frac{1}{\kappa} \sqrt{f(r)}  e^{-k r_*(r)}
\end{equation}
In this formalism, the geodesic distance between two points (a,b) is given by -
\begin{equation}
    L(a,b) = \Omega(a) \Omega(b) \big[ U(b) - U(a) \big]\big[ V(a) - V(b) \big] \label{eq13}
\end{equation}

Furthermore, we consider the bath is in Minkowski spacetime which is characterized by the metric $ds^2 = -dt^2 + dr^2 $. Thus the conformal factor for the flat bath region is given by - 
\begin{equation}
     \Omega (r) = \frac{1}{\kappa}   e^{-k r}
\end{equation}
 \section{PAGE CURVE FOR NON EXTREMAL KERR ADS BLACK HOLE}\label{sec3}
 In this section, we calculate the entanglement entropy for 4D Kerr AdS black hole. We describe two secenero, one without island and the other with one island by using the formula (\ref{eq1}) i.e. -
\begin{equation}
    S_\text{gen} = \text{min}\{ext[ \frac{Area (\partial I)}{4 G_N} + S_\text{field} (R \cup I)]\} \label{eq5} 
\end{equation}
 To calculate the entanglement entropy, we employ the large distance limit and s-wave approximation. In this way we neglect the spherical part of the metric and consider only the conformally flat part.   This assumption is based on the fact that Hawking radiation observed by a distant observer can be described accurately with 2D CFT as derived in \cite{hashimoto2020islands}.
 
 \subsection{ Without Island}

 \par At first, we calculate the entanglement entropy without considering an island.  The Penrose diagram for non extremal Kerr AdS black hole is shown in figure. The radiation is collected in the region $R_{\pm}$ and this region is far from the horizon of the black hole. The boundary point of the radiation region is given by $b_{\pm}$. We assume the whole system is in pure state at t=0. We need to calculate the entropy of the region outside $(b_{+}, b_{-})$ which is equivalent to the entropy of the region $(b_{+}, b_{-})$ according to the complementary principle of Von Neumann entropy. The coordinates are $(t,r_b) = (t_{b},r_b)$  for $(b_{+}) $ and $(t,r_b) = (-t_{b}+\frac{i\beta}{2},r_b)$  for $(b_{-}) $ respectively. The entropy of the CFT radiation is given by -
 \begin{equation}
     S_{Bulk} = S(\text{R})= \frac{c}{6} \log(d(b_{+}, b_{-}) )\nonumber
 \end{equation}
 where $d(b_{+}, b_{-})$ is the geodesic distance given by (\ref{eq13}) between the points $(b_{+}, b_{-})$ and c is the central charge. Therefore, the entanglement entropy without island is given by-
 \begin{align}
      S(R) = \frac{c}{6} \log\bigg[\frac{4 \cosh^2(k t_{b})}{\kappa^2}\bigg]
 \end{align}
 At late times ( $t_b \rightarrow{\infty}$), the above equation can be written as -
 \begin{align}
     S(R) &\simeq \frac{c}{6} \log\bigg[\frac{e^{2 \kappa t_{b}}}{\kappa^2}\bigg] \\ &\simeq \frac{c}{3} k t_b \label{eq18}
 \end{align}
From the above equation it is clear that the entanglement entropy of the radiation grows linearly with time and no Page time shows up in this calculation. This shows that at late times it will  eventually exceed the Hawking Bakenstein entropy of the black hole that leads to the loss of information of the black hole. For an eternal black hole, the entanglement entropy can be at most twice the Hawking Bekenstein entropy\cite{bekenstein1981universal} where as the above equation violates the same.  This makes the paradox more pronounced. In the subsequent calculations, we will work on the existence of one island scenario. 
 
     \begin{figure}[t]
    \centering
    \begin{subfigure}[b]{0.45\textwidth}
        \centering
        \includegraphics[width=\textwidth]{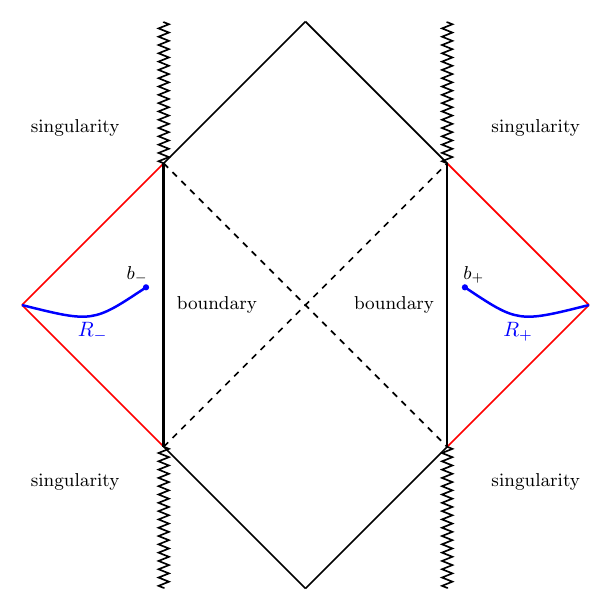}
        \caption{}
        \label{f1a}
    \end{subfigure}
    \hspace{0.03\textwidth}
    \begin{subfigure}[b]{0.45\textwidth}
        \centering
        \includegraphics[width=\textwidth]{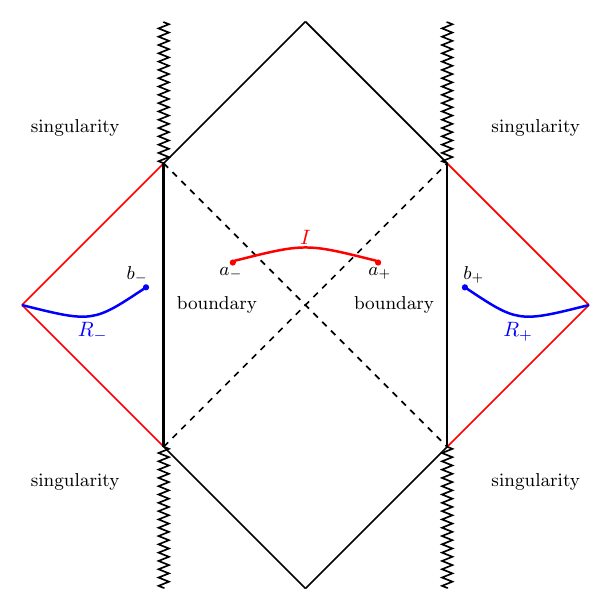}
        \caption{}
        \label{f1b}
    \end{subfigure}
    {\caption{(a) The Penrose diagram of an eternal Kerr $\text{AdS}_4$ black hole without island.The black and red regions denote the black hole and the beth region respectively. The radiation regions are denoted by $R_{\pm}$ on the right and left wedges and $b_{\pm}$ are the boundaries of $R_{\pm}$ (b) The Penrose diagram of an eternal Kerr $\text{AdS}_4$ black hole with an island. The boundary of the island is denoted as $a_{\pm}$  The radiation regions are denoted by $R_{\pm}$ on the right and left wedges and $b_{\pm}$ are the boundaries of $R_{\pm}$.}}
    \label{Penrose_island}
\end{figure}

 \subsection{ With Island}
In this section we calculate the entanglement entropy by considering contributions from one island with boundaries $a_{+}$  and $ a_{-}$. They are located at $(t,r_a) = ( t_a, r_a)$ for $a_{+}$ and $(t,r_a) = ( -t_a-\frac{i \beta}{2}, r_a)$ for $a_{-}$. Following equation (\ref{eq1}), the entanglement entropy of the field in the region $R \cup I$ is given by- 
\begin{equation}
    S_\text{field} (R \cup I) = \frac{c}{6}
 \log\bigg[ \frac{L(a_{+}, a_{-}) L(b_{+}, b_{-}) L(a_{+}, b_{+}) L(a_{-}, b_{-}) }{L(a_{+}, b_{-}) L(a_{-}, b_{+})}\bigg]
 \end{equation}
The first term (area term) in equation (\ref{eq1}) is given by \begin{equation}
   \frac{Area (\partial I)}{4 G} = \frac{4 \pi (r^2_{a}+a^2)}{4 G \Xi} = \frac{\pi (r^2_{a}+a^2)}{ G \Xi} 
\end{equation}
Then, the generalised entropy can be written as -
\begin{align}
     S_\text{gen} = \frac{2 \pi (r^2_{a}+a^2)}{ G \Xi} + \frac{c}{6} \log\bigg[ \frac{16 f(r_a) \cosh^2(\kappa t_a) \cosh^2(\kappa t_b) }{\kappa^4}\bigg] \nonumber \\
 + \frac{c}{3} \log\bigg[\frac{\cosh[\kappa(r_*(r_a)-r_*(r_b)]-\cosh[\kappa(t_a - t_b)]}{\cosh[\kappa(r_*(r_a)-r_*(r_b)]+\cosh[\kappa(t_a + t_b)]} \bigg] \label{eq21}
\end{align}
\textbf{At early times} we assume the  approximations as-
\begin{equation}
    t_a, t_b \simeq 0, \qquad t_a, t_b \ll r_{+}, \qquad t_a, t_b \ll \frac{1}{\kappa} \ll r_{*}(r_{b})-r_*(r_a),\qquad b\gg r_+
\end{equation}
   As we pick the cut off surface far away from the horizon, the generalised entropy given by equation (\ref{eq21}) can be approximated as -
   \begin{align}
       S_\text{gen}^{early} &\simeq \frac{2 \pi (r^2_{a}+a^2)}{ G \Xi} + \frac{c}{6} \log\bigg[ \frac{16 f(r_a) \cosh^2(\kappa t_a) \cosh^2(\kappa t_b) }{\kappa^4}\bigg] 
   \end{align}
   By extremising the above equation with respect to $t_a$ and $r_a$ we get the following two equations-
   \begin{align}
       \frac{\partial  S_\text{gen}^{early}}{\partial t_a} &= \frac{c}{3} \tanh(\kappa t_a) =0 , \label{eq24}\\
       \frac{\partial  S_\text{gen}^{early}}{\partial r_a} &= \frac{4 \pi r_a}{ G \Xi} + \frac{c}{6} \frac{f^{'}(r_a)}{f^(r_a)} = 0.\nonumber \\ 
       &= \frac{4 \pi r_a}{ G \Xi} + \frac{c}{6}\frac{a^2 (2r_a -r_{-}-r_{+})-r_a (2 r_{+}r_{-}-r_a(r_{+}+r_{-})}{ (r_a^2 + a^2)(r_a-r_{+})(r_a-r_{-})}=0\label{eq25}
   \end{align}
Solving equation (\ref{eq24}) we get $t_a = 0$. From equation (\ref{eq25}), neglecting the higher order terms, the location of the island is given by
\begin{align}
   r_a &\simeq \frac{c G(r_+ + r_-) \Xi}{24 \pi r_+ r_-} \nonumber \\
   &\simeq \frac{c (r_+ + r_-) \Xi}{24 \pi r_+ r_-} l_p^2
\end{align}
where $\sqrt{G} = l_p$ is the plank length and we should ignore the physics in the length length regime. This shows that, no island is formed at the early times and the value of the entanglement entropy is determined by equation (\ref{eq18}). \\ 
\textbf{At late times}, we calculate the entanglement entropy with the following approximation:  
     \begin{equation}
\begin{aligned}
&t_a,\, t_b \gg b > r_{+},
\qquad 
\cosh\!\bigl(\kappa(t_a+t_b)\bigr)
\gg 
\cosh\!\bigl(\kappa(r_*(r_a)-r_*(r_b))\bigr),\\[6pt]
&\cosh\!\bigl(\kappa(r_*(r_a)-r_*(r_b))\bigr)
\simeq 
\frac{1}{2}\exp\!\bigl[\kappa\,(r_*(r_b)-r_*(r_a))\bigr].
\end{aligned}
\end{equation}
Furthermore, as the distance between left wedge and  right wedge is very large at the late time scale ( $(t_a, t_b \gg b)$, following \cite{hashimoto2020islands} we can have the following approximations -
\begin{equation}
    L(a_+,a_-) \simeq L(b_+,b_-) \simeq L(a_+,b_-) \simeq L(a_-, b_+) \gg  L(b_{\pm},b_{\mp})  
\end{equation}
Plugging these approximation into equation (\ref{eq21}), the generalised entropy at late times is given by 
\begin{equation}
     S_\text{gen} = \frac{2 \pi (r^2_{a}+a^2)}{ G \Xi} + \frac{c}{6} \log\bigg[ \frac{4 {f(r_a)}  }{\kappa^4}  
 \bigg(\cosh[\kappa(r_*(r_a)-r_*(r_b)]-\cosh[\kappa(t_a - t_b)]\bigg)^2 \bigg] \label{eq29}
\end{equation}

\begin{figure}[t]
    \centering
        \includegraphics[width=0.45\textwidth]{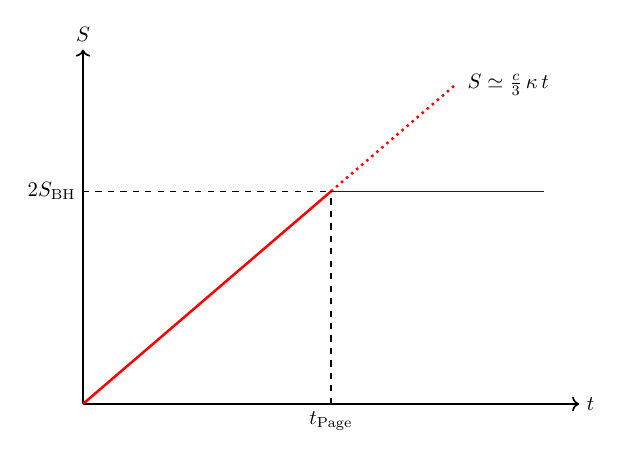}
        \caption{The Page curve of a non extremal Kerr $\text{AdS}_4$ black hole represented by the solid line. The red line and blue line represents the entanglement entropy without and with island respectively. }

    \label{PageCurve}
\end{figure}

Following the previous method, by extremising the above equation  with respect to $t_a$ we get -
\begin{equation}
    \frac{\partial S_{\text{gen}}^{late}}{\partial t_{a}} = \frac{c}{3} \frac{\kappa \sinh(\kappa(t_a - t_b))}{\cosh[\kappa(r_*(r_a)-r_*(r_b)]-\cosh[\kappa(t_a - t_b)} = 0
\end{equation}
It is clear that we above equation is satisfied only when $t_a = t_b$. By putting $t_a = t_b$ in equation (\ref{eq29}), we get -
\begin{align}
    S_{\text{gen}}^{late} &=  \frac{2 \pi (r^2_{a}+a^2)}{ G \Xi} + \frac{c}{3}\log\bigg[\frac{2 \sqrt{{f(r_a)}}}{\kappa^2}  
 \bigg(\cosh[\kappa(r_*(r_a)-r_*(r_b)]-1 \bigg)\bigg] \nonumber \\ 
 &= \frac{2 \pi (r^2_{a}+a^2)}{ G \Xi} + \frac{c}{3}\log\bigg[\frac{2 \sqrt{f(r_a)}}{\kappa^2}\bigg] + \frac{c \kappa}{3} \big(r_*(r_b)-r_*(r_a)) - \frac{2 c}{3} e^{-\kappa(r_*(r_b)-r_*(r_a))} 
 \end{align}
Extremising the above equation with respect to $r_a$, we find 
\begin{equation}
    \frac{\partial S_{\text{gen}}^{late}}{\partial r_{a}} = \frac{4 \pi r_{a}}{ G \Xi} + \frac{c}{6} \frac{f'(r_a)}{f(r_a)} - \frac{ c \kappa }{3 f(r_a)}\big(1+2 e^{-\kappa(r_*(r_b)-r_*(r_a)} \big) = 0 \label{eq32}
\end{equation}
At late times and large distance limit we utilize the near horizon limit i.e. $r_a \simeq r_{+}$ and approximate the tortoise co-ordinate as follows -
\begin{align}
   f(r) &\simeq f^{'} (r)(r-r_{+}) + \mathcal{O}(r-r_+)^2 
 =  2\kappa (r -r_{+}) ; 
 \end{align}
  \begin{align}
     r_*(r) &= \int \frac{1}{f(r)} dr \nonumber  \\
     &= \frac{1}{2 \kappa} \log |\frac{r-r_+}{r_+}| 
     \label{eq34}
  \end{align}
   By substituting the above approximations back into equation (\ref{eq32}) we find the location of the island as -
   \begin{equation}
       r_a \simeq r_+ + \frac{c^2 G^2 \Xi^2 }{144 \pi^2 r_{+}^3} e^{-2 \kappa b} \label{eq35}
   \end{equation}
Thus, we get the entanglement entropy at late times as -
\begin{equation}
    S = \frac{2 \pi (r^2_{+}+a^2)}{ G \Xi} + \mathcal{O}(cG) = 2 S_{BH} \label{eq36}
\end{equation}
   Thus, the entanglement entropy at late time is dominated by the area term, which is a constant and does not grow with time unlike the early times where entanglement entropy grows monotonically with time. Therefore, when entanglement islands are taken into account, the entanglement entropy at late times approaches a constant value equal to twice the Bekenstein–Hawking entropy of an eternal Kerr-AdS black hole which is consistent with Hawking Bekenstein entropy bound. This saturation ensures consistency with unitary evolution and indicates the information preservation  in the  non-extremal Kerr–AdS black hole. This is consistent with the expected Page curve behavior as illustrated in Fig. (\ref{PageCurve})

\subsection{Page time and scrambling time}
In this section we calculate the Page time and Scrambling time. Page time is the time when the entanglement entropy reaches its maximum value. In other words, Page time is the point of intersection of entropy curve as shown in Fig (\ref{PageCurve}).  The value of entanglement entropy at early times is given by equation \ref{eq18} whereas at late times the entanglement entropy is given by (\ref{eq36}). By equating (\ref{eq18}) and (\ref{eq36})  We get the value of Page time as - 
\begin{align}
  t_{Page} &= \frac{6}{c \kappa} S_{\text{BH}} \nonumber \\
&=   \frac{3}{c \pi T(r_+)} S_{\text{BH}}
\end{align}
Further, we comment on the scrambling time. It  is the shortest time required to recover information by an outside observer via Hawking radiation that previously fell into the black hole\cite{hayden2007black}. Using equations \ref{eq34} and \ref{eq35}, the scrambling time is given by\cite{Lin:2024gip} 
\begin{equation}
    t_{scr} \equiv Min[\Delta t ] = r_*(r_b)-r_*(r_a) \nonumber
\end{equation}
\begin{align}
    t_{scr} &= b- \frac{1}{2 \kappa} \log\big| \frac{c^2 G^2 \Xi^2}{144 \pi^2 r_{+}^4 }e^{-2 \kappa b}\big|\nonumber  \\ &\simeq \frac{1}{\kappa} \log\big| \frac{12 \pi r_{+}^2 }{cG\Xi} e^{ \kappa b}\big| \nonumber  \\ 
    &\simeq \frac{1}{k} \log S_{BH}
    \end{align}
    Where we used equation (\ref{eq35}), near horizon limit. Also, $r{_+} \gg$ than other terms.


\section{Phase transitions and page curve}\label{sec4}
In this section we investigate emergence of first-order phase transitions in kerr-AdS black holes and assess their possible impact on the entanglement entropy of an evaporating black hole.We divide our analysis into two parts. First, we investigate the thermodynamic behaviour in the canonical ensemble. Subsequently, we introduce an alternative ensemble in which a new thermodynamic variable $\zeta$ conjugate to the spin parameter $a$ is defined. Within this framework, we examine the phase structure and analyze the behaviour of the entanglement entropy.To investigate the phase structure of the Kerr–AdS black hole, we employ the standard thermodynamic approach of computing the free energy and examining its dependence on the Hawking temperature. The presence of a first-order phase transition is signaled by the emergence of a characteristic swallow-tail structure in the free-energy curve.

\subsection{Canonical Ensemble}
The free energy of the Kerr–AdS black hole in the canonical ensemble is obtained using
\begin{equation}\label{eqfree}
    F=M-TS
\end{equation}
where $M$, $T$ and $S$ are the mass, temperature and entropy of the black hole in the canonical ensemble. The explicit expressions are given below respectively
\begin{equation}\label{eqmass}
    M=\frac{1}{2 \left(1-\frac{a^2}{l^2}\right)^2 r_+}\left(\frac{a^2 r_+^2}{l^2}+\frac{r_+^4}{l^2}+a^2+r_+^2\right)
\end{equation}
\begin{equation}\label{eqtemp}
    T=\frac{r_+}{4 \pi  \left(a^2+r_+^2\right)}\left(\frac{a^2}{l^2}-\frac{a^2}{r_+^2}+\frac{3 r_+^2}{l^2}+1\right)
\end{equation}
\begin{equation}\label{eqentropy}
    S=\frac{\pi  \left(a^2+r_+^2\right)}{1-a^2 l^{-2}}
\end{equation}

\begin{figure}[h!]
    \centering
    \begin{subfigure}[b]{0.45\textwidth}
        \centering
        \includegraphics[width=\textwidth]{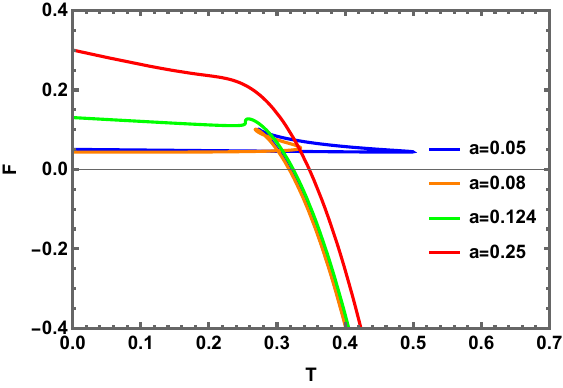}
        \caption{}
        \label{f1a}
    \end{subfigure}
    \hspace{0.03\textwidth}
    \begin{subfigure}[b]{0.45\textwidth}
        \centering
        \includegraphics[width=\textwidth]{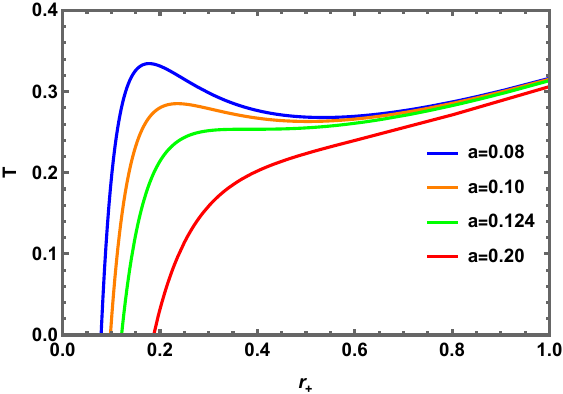}
        \caption{}
        \label{f1b}
    \end{subfigure}

    \caption{(a) The plot of free energy versus temperature showing the swallow-tail behaviour for below critical values of the spin parameter. (b) The variation of the Hawking temperature with the horizon radius. Below the critical temperature, three black hole branches are observed.}
    \label{t v r}
\end{figure}
In Fig. \ref{t v r}, we present the free energy as a function of temperature and the Hawking temperature as a function of the horizon radius. We consider the AdS length to be $1$ throughout the analysis. The left panel clearly demonstrates the emergence of a swallow-tail structure when the spin parameter $a$ is below its critical value of $a_c=0.12455$. 
The critical values are obtained by solving the equations 
\begin{equation}\label{eqcritical}
    \frac{dT}{dr}=0, \quad \frac{d^2T}{dr^2}=0
\end{equation}
The swallow-tail behavior is a hallmark of a first-order phase transition. For values of the spin parameter $a$ below its critical value, the free energy exhibits three distinct branches, as illustrated by the blue and orange curves in Fig. \ref{f1a}. These correspond to the small black hole (SBH), intermediate black hole (IBH), and large black hole (LBH) phases. Among them, the IBH branch possesses the highest free energy and is therefore thermodynamically unstable. The point of intersection of the swallow is termed as the phase transition point where a first-order phase transition occurs between the small and the large black hole phase. When the spin parameter exceeds its critical value, these three branches merge and the swallow-tail structure disappears, yielding a smooth, continuous free-energy curve, as shown by the red curve in Fig. \ref{f1a}.

In the right panel, we display the variation of the Hawking temperature with the horizon radius $r_+$. For values of the spin parameter below the critical value, $a<a_c$, the black hole exhibits three distinct branches, as indicated by the blue and orange curves in Fig.~\ref{f1b}. In this regime, a given Hawking temperature corresponds to three possible values of the horizon radius, signaling the presence of small, intermediate, and large black hole phases. When the spin parameter exceeds the critical value, $a>a_c$, the Hawking temperature becomes a single-valued function of $r_+$ and increases monotonically, indicating the disappearance of the multiple-branch structure. For illustration, at $a=0.08<a_c$, the small black hole branch extends up to $r_+=0.1775$, the intermediate black hole branch lies in the range $0.1775<r_+<0.5350$, and the large black hole branch begins at $r_+=0.5350$.

We now proceed to examine the relationship between the Page time and phase transitions in Kerr–AdS black holes. We anticipate that the occurrence of a phase transition, whether before or after the Page time, may significantly influence the behavior of the Page curve. In order to investigate the phase transitions, we must restrict the spin parameter as $a<a_c$.

The entanglement entropy is described as follows
\begin{equation}\label{eq_entgl}
     S = \text{Min}{ \left [ \frac{2}{3} \pi T(r_{+}) t,\ 2 S_{BH}  \right ] }
\end{equation}
Furthermore, the Page time exhibits temperature dependence. In the regime where the horizon radius $r_+$ grows linearly, the Page time may be approximated by the relation below
\begin{equation}\label{eq_page}
     t_{\text{Page}}  \simeq \frac{3S_{BH}}{c \pi T(r_+)}
\end{equation}
\begin{figure}[ht]
    \centering
    \begin{subfigure}{0.32\textwidth}
        \centering
        \includegraphics[width=\linewidth]{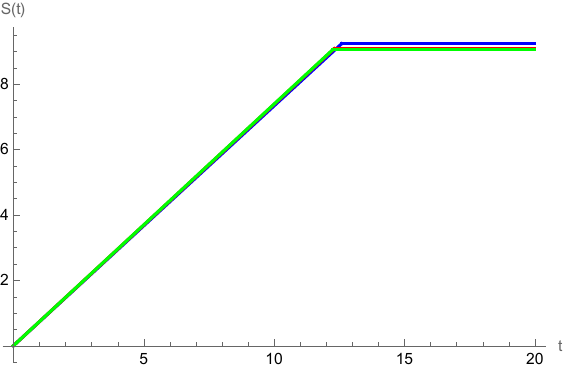}
        \caption{}
        \label{fig:a}
    \end{subfigure}
    \hfill
    \begin{subfigure}{0.32\textwidth}
        \centering
        \includegraphics[width=\linewidth]{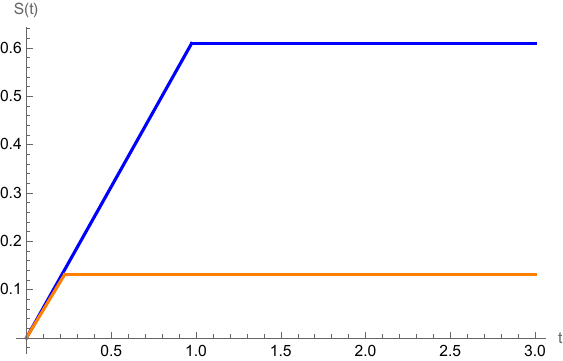}
        \caption{}
        \label{fig:b}
    \end{subfigure}
    \hfill
    \begin{subfigure}{0.32\textwidth}
        \centering
        \includegraphics[width=\linewidth]{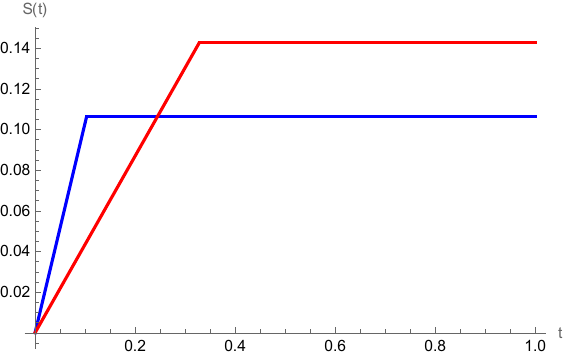}
        \caption{}
        \label{fig:c}
    \end{subfigure}
    \caption{The Page curves of eternal Kerr AdS black holes. (a) Page curves for spin parameter $a=0.11$ (blue), $a=0.05$ (red) and $a=0.01$ (green) for large black hole. (b) Page curves for $a=0.08$ with blue line for intermediate black hole and orange line for small black hole. (c) Page curve for small black hole $r_+=0.12$, $a=0.05$ (blue) and $a=0.09$ (red)}
    \label{fpage}
\end{figure}
We present the Page curves for Kerr–AdS black holes in Figure \ref{fpage}. Figure \ref{fig:a} displays the Page curves for different values of the spin parameter $a$ in the large black hole regime ($r_+=1.2$). We observe that the Page time shifts as the spin parameter varies. In particular, decreasing $a$ causes the Page time to occur earlier. Figure \ref{fig:b} displays the Page curves for the intermediate and small black hole phases at a fixed spin parameter $a=0.08$. The blue curve corresponds to the intermediate black hole phase ($r_+=0.3$), while the orange curve represents the small black hole phase ($r_+=0.12$). We observe that, for a given value of the spin parameter, the Page time of the small black hole occurs earlier than that of the intermediate black hole. Finally we compare the Page curves for the small black hole phase ($r_+=0.12$) for different spin parameters in Figure \ref{fig:c}. We can again observe here that as the spin parameter is decreased, the Page time also occurs at an early stage. From these observations, we conclude that the spin parameter significantly influences both the phase transitions and the Page time.

Next, we adopt a framework motivated by earlier studies \cite{almheiri2019entropy,Rocha_2008,geng2025openadscftdoubletrace,Geng_2025}. We consider a configuration in which one boundary is connected to an external auxiliary system that functions as an energy reservoir. This interaction enables a two-sided black hole to evaporate through one side. When such a coupling is present, the black hole continuously exchanges energy with its environment, leading to a slow reduction of its mass. As the evaporation progresses, both the temperature and entropy of the black hole evolve, and these variations are manifested in the structure of the Page curve. If the external baths are removed, the black hole becomes effectively isolated and follows a qualitatively different evaporation path. Without access to an external energy reservoir, the black hole’s temperature increases rapidly, which in turn accelerates the evaporation process. This abrupt evolution can imprint distinctive features on the Page curve, such as a sharper descent or possible discontinuities. We now incorporate these effects into our analysis.

In what follows, we study the structure of the Page curve in different ranges of the event horizon radius, namely the horizon radii corresponding to small, large and intermediate black holes. During black hole evaporation, the horizon radius $r_+$ becomes time-dependent. Following \cite{Page_2005,Konoplya_2019}, its evolution obeys $\frac{dr_+}{dt} = -x$, where $x$ is a constant determined by fundamental constants including $\hbar$, $c$, $G$, and the mass $M$ of the black hole. We set the value of $a$ as $0.08$ which is way below its critical value and hence confirming a phase transition.

\begin{figure}[ht]
    \centering
    \begin{subfigure}{0.31\textwidth}
        \centering
        \includegraphics[width=\linewidth]{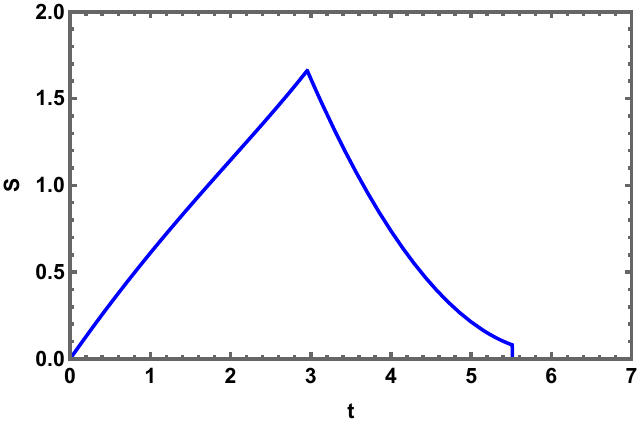}
        \caption{}
        \label{fig:a2}
    \end{subfigure}
    \hfill
    \begin{subfigure}{0.32\textwidth}
        \centering
        \includegraphics[width=\linewidth]{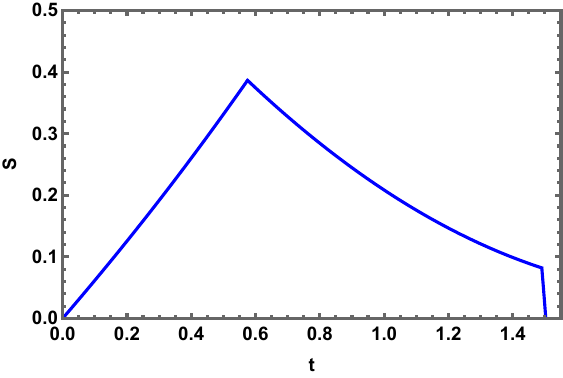}
        \caption{}
        \label{fig:b2}
    \end{subfigure}
    \hfill
    \begin{subfigure}{0.33\textwidth}
        \centering
        \includegraphics[width=\linewidth]{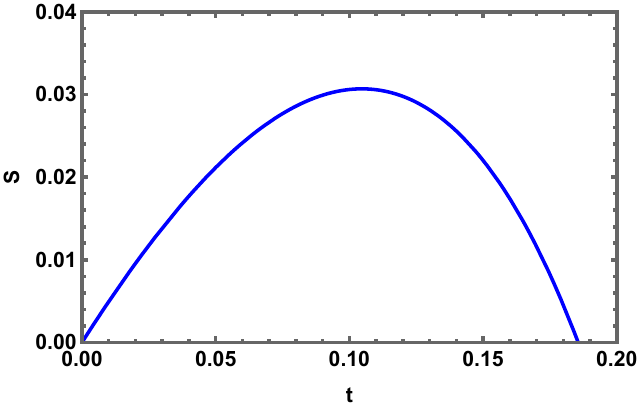}
        \caption{}
        \label{fig:c2}
    \end{subfigure}
    \caption{The Page curve for a Kerr-AdS black hole that is initially coupled to an external bath and subsequently isolated to undergo evaporation. (a) Page curve for Kerr AdS black hole in the large black hole regime. (b) Page curve of a Kerr AdS black hole with an intermediate horizon radius. (c) Page curve of for a small Kerr-AdS black hole.}
    \label{fpage5}
\end{figure}

We show the Page curves in Figure \ref{fpage5}. Figure \ref{fig:a2} depicts the Page curve corresponding Kerr-AdS black hole having a horizon radius in the large black hole regime. We observe that at early times, the Page curve exhibits a monotonic increase; however, after a time slightly exceeding half of the evaporation duration, the curve exhibits a downward trend and ultimately the entropy approaches zero. There is a sharp discontinuity in the Page curve before the entropy reaches zero. Assuming that the horizon radius shrinks linearly with time, the entanglement entropy attains a minimum value, and the Bekenstein–Hawking entropy contribution can be represented as a piecewise function composed of three distinct segments. This structure gives rise to a Page curve that incorporates the imprint of the Hawking–Page phase transition, where the observed discontinuity originates directly from the phase transition \cite{Lin:2024gip}.

Figure \ref{fig:b2} depicts the Page curve for Kerr-AdS black hole with an intermediate horizon radius. Initially, the Page curve is observed to increase linearly. After the Page time, the curve takes a downward trend and finally reaches zero. The peak of the entropy for the intermediate black hole case is lower than that of the larger black hole case. The Page time also occurs at an earlier stage. Before entropy goes to zero, the Page curve exhibits a sharper discontinuity compared to the corresponding curve for a large black hole. The intermediate branch corresponds to the region of free energy with negative slope and is known to be unstable in the phase structure of Kerr–AdS black holes. Consequently, the distinctive shape of the Page curve in this regime encodes the presence of the first-order phase transition between the small and large black hole phases.

The Page curve for the small Kerr–AdS black hole (Figure \ref{fig:c2}) exhibits the standard unitary behavior expected for an evaporating black hole. The entanglement entropy initially increases with time, reflecting the accumulation of entanglement between the black hole and the emitted Hawking radiation. It reaches a maximum at the Page time, after which the entropy decreases monotonically and eventually vanishes as the black hole completes its evaporation. At low temperatures, only a single small black hole branch exists; consequently, no phase transition occurs and the Page curve exhibits no discontinuity \cite{Lin:2024gip}.

\subsection{Fixed $\zeta$ Ensemble}
\begin{figure}[h!]
    \centering
    \begin{subfigure}[b]{0.45\textwidth}
        \centering
        \includegraphics[width=\textwidth]{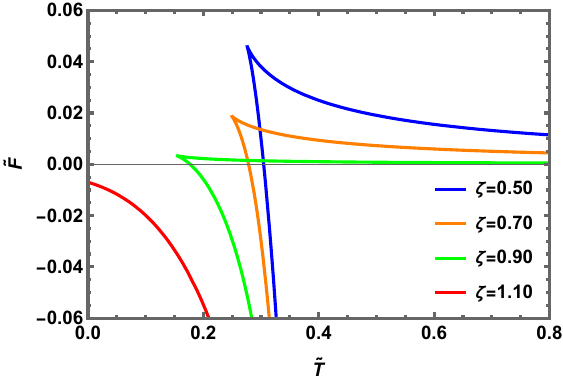}
        \caption{}
        \label{f1afixed}
    \end{subfigure}
    \hspace{0.03\textwidth}
    \begin{subfigure}[b]{0.45\textwidth}
        \centering
        \includegraphics[width=\textwidth]{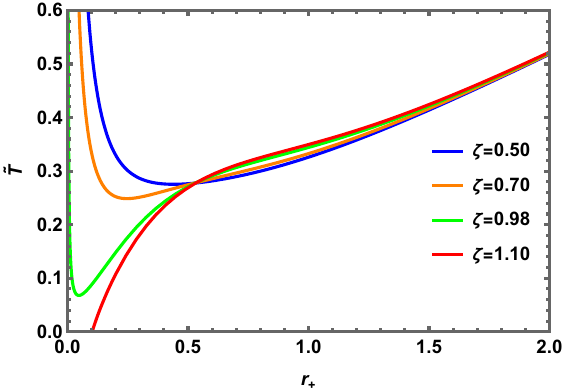}
        \caption{}
        \label{f1bfixed}
    \end{subfigure}

    \caption{(a) The plot of free energy versus temperature showing the the two black hole branch for below critical values of the $\zeta$. (b) The variation of the Hawking temperature with the horizon radius. Below the critical temperature, two black hole branches are observed.}
    \label{t v r fixed}
\end{figure}

\begin{figure}[ht]
    \centering
    \begin{subfigure}{0.32\textwidth}
        \centering
        \includegraphics[width=\linewidth]{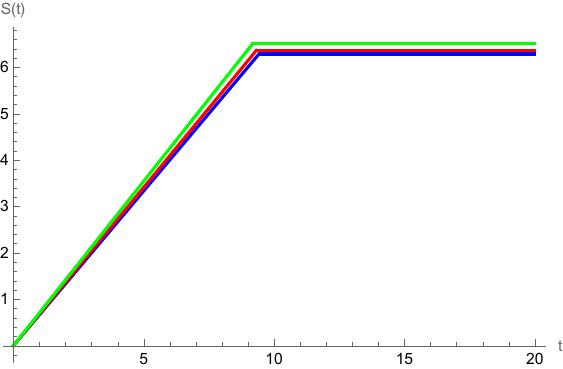}
        \caption{}
        \label{fig:ax}
    \end{subfigure}
    \hfill
    \begin{subfigure}{0.32\textwidth}
        \centering
        \includegraphics[width=\linewidth]{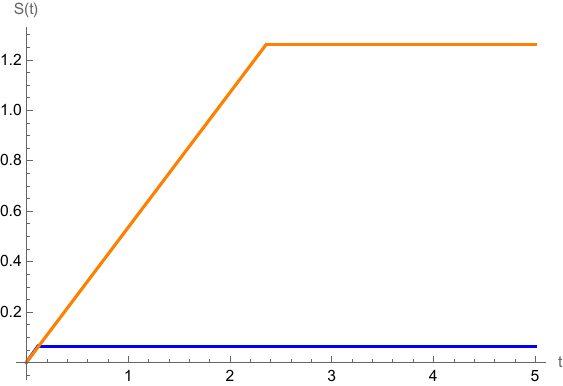}
        \caption{}
        \label{fig:bx}
    \end{subfigure}
    \hfill
    \begin{subfigure}{0.32\textwidth}
        \centering
        \includegraphics[width=\linewidth]{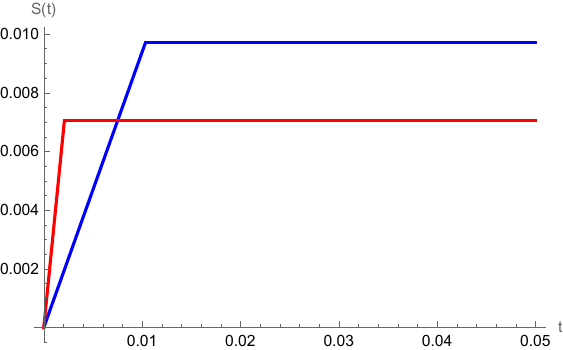}
        \caption{}
        \label{fig:cx}
    \end{subfigure}
    \caption{The Page curves of eternal Kerr AdS black holes. (a) Page curves for $\zeta=0.1$ (blue), $\zeta=0.5$ (red) and $\zeta=0.85$ (green) for large black hole. (b) Page curves for $\zeta=0.8$ with blue line for small black hole and orange line for large black hole. (c) Page curve for small black hole $r_+=0.03$, $\zeta=0.85$ (blue) and $\zeta=0.5$ (red)}
    \label{fxfpage}
\end{figure}

In this subsection, we consider a new ensemble where we take a new quantity $\zeta$ which is conjugate to the spin parameter $a$. It is defined as $\zeta=\frac{dM}{da}$. Taking the derivative we get
\begin{equation}\label{eq_zeta}
    \zeta=\frac{a \left(r^2+1\right) \left(a^2+2 r^2+1\right)}{\left(a^2-1\right)^3 G r}
\end{equation}
The mass in the new ensemble can be found out by taking the Legendre transformation $\Tilde{M}=M-\zeta a$. The entropy ($\Tilde{S}$) can be written by expressing $a$ in terms of $\zeta$ in equation \ref{eqentropy}, and the temperature can be easily computed using $\Tilde{T}=\frac{d\Tilde{M}}{d\Tilde{S}}$. Therefore, the free energy can be expressed using $\Tilde{F}=\Tilde{M}-\Tilde{T}\Tilde{S}$ The expression are large and cumbersome so we avoid writing them explicitly. We show the plot of the free energy with temperature and the variation of the Hawking temperature with the horizon radius in Figure \ref{t v r fixed}.
We observe that in the new ensemble, we observe only two distinct branches of the black hole (Figure \ref{f1afixed}), namely the small and the large black hole branch, in contrast to the three branches in the canonical ensemble. This should not be surprising as the thermodynamics of black hole is highly ensemble dependent. From Fig. \ref{f1afixed}, we can observe that there is no swallow-tail behaviour, and therefore the first order phase transition is absent in this ensemble and so no criticality exists- the phase transition is a Hawking-Page transition in this case. However, we can confirm that there are two black hole branches when the parameter $\zeta$ is equal to or below, $0.90$ (the green, orange and blue curves) and for $\zeta>1.10$, the free energy becomes a smooth continuous curve and only one branch shown by the red line remains. We now proceed to examine the relationship between the Page time and phase transitions in the new ensemble. We take care of the fact that the value of $\zeta$ is always smaller than $0.90$ in order to ensure a phase transition. We show the Page curves for Kerr-AdS black hole in the fixed $\zeta$ ensemble in Figure \ref{fxfpage}.

Figure \ref{fig:ax} depicts the Page curve for different values of the parameter $\zeta$ in the large black hole regime of the Kerr black hole. The Page time is seen to depend on $\zeta$, particularly, decreasing the value of $\zeta$ shifts the Page time to to occur at an earlier stage. In Figure \ref{fig:bx} we show the Page curves for the small and the large black hole phases in the fixed $\zeta=0.8$ ensemble. In this ensemble as well, we find that the Page time of the small black hole occurs earlier than that of the large black hole. We further compare the Page curves of the small black hole phase for different values of $\zeta$ in Figure \ref{fig:cx}. Similar to the canonical ensemble, lowering the parameter $\zeta$ (or $a$ in the canonical ensemble) shifts the Page time to an earlier stage.

\begin{figure}[h!]
    \centering
    \begin{subfigure}[b]{0.45\textwidth}
        \centering
        \includegraphics[width=\textwidth]{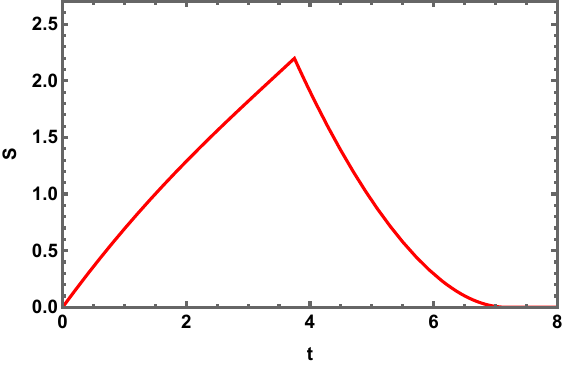}
        \caption{}
        \label{page_z1}
    \end{subfigure}
    \hspace{0.03\textwidth}
    \begin{subfigure}[b]{0.45\textwidth}
        \centering
        \includegraphics[width=\textwidth]{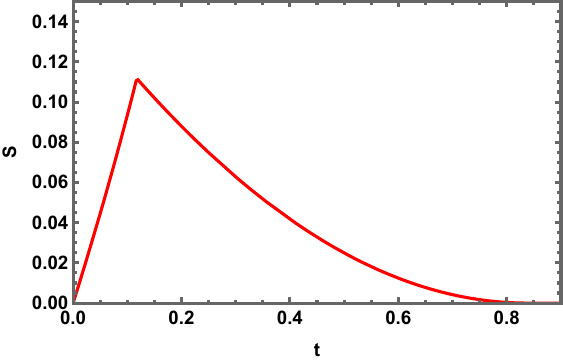}
        \caption{}
        \label{page_z2}
    \end{subfigure}

    \caption{The Page curve for a Kerr-AdS black hole in fixed $\zeta$ ensemble that is initially coupled to an external bath and then isolated to undergo evaporation. (a) Page curve in the large black hole regime. (c) Page curve of for a small Kerr-AdS black hole.}
    \label{page_z}
\end{figure}

Next we proceed to investigate the Page curves for the Kerr-AdS black hole in fixed $\zeta$ ensemble which is initially coupled to an external bath and then isolated to undergo evaporation. The computations are the same as discussed in the previous subsection. We present the Page curves of such a black hole in the fixed $\zeta$ ensemble in Figure \ref{page_z}.
Figure \ref{page_z1} shows the Page curve for the large black hole phase in the fixed–$\zeta$ ensemble. The entanglement entropy initially increases linearly with time, reflecting the gradual build-up of entanglement between the black hole and the emitted Hawking radiation. The curve reaches a maximum at the Page time, after which the entropy begins to decrease, signaling the onset of information release. Eventually, the entropy approaches zero as the black hole completes its evaporation. Compared to the canonical ensemble, the location of the Page time is shifted due to the presence of the fixed–$\zeta$ constraint, demonstrating that the ensemble choice influences the evaporation dynamics. Nevertheless, the qualitative behavior remains consistent. Moreover, since no first-order phase transition occurs, the Page curve remains continuous and does not exhibit any discontinuity.

Figure~\ref{page_z2} displays the Page curve for the small black hole phase in the fixed--$\zeta$ ensemble. The entanglement entropy rises rapidly at early times, indicating a swift accumulation of entanglement between the black hole and the Hawking radiation. The curve attains its maximum at an early Page time, after which the entropy decreases monotonically and eventually vanishes as the black hole evaporates completely. Compared to the large black hole phase, the Page time occurs significantly earlier and the peak entropy is lower, reflecting the smaller horizon radius and reduced thermodynamic capacity of the small black hole. This behavior further confirms that, in the fixed-$\zeta$ ensemble, the small black hole evaporates more rapidly and releases information at an earlier stage.

\section{discussion and conclusion}\label{sec5}
In this paper we investigate the information paradox in light of entanglement entropy for Kerr AdS black hole. At first we utilize a dimensional reduction technique to obtain an effective 2D field theory and then we calculate the entanglement entropy of radiation within the two dimensional framework. We consider two scenarios, one with island and other without island. In the first case, we found that the entanglement entropy of the Hawking radiation increases monotonically with time as described by the equation (\ref{eq18}) and no Page time exists. This contradicts the black hole's thermodynamics entropy bound and violates the unitary principle of quantum mechanics. After considering the contributions from the entanglement island, we found that, at late times the entropy of the Hawking radiation reaches constant value and that does not violate the entropy bound. This saturation ensures consistency with unitary evolution and indicates the information preservation in the  Kerr AdS black hole. Based on these findings, we plot the the Page curve as shown in Fig (\ref{PageCurve}) which is consistent with the unitary principle of quantum mechanics. We then calculate the Page time and Scrambling time.
\par Next, we investigate the influence of thermodynamic phase transitions on the Page curves of Kerr-AdS black holes in two different ensembles. We first analyze the first-order phase transition in the canonical ensemble and present the Page curves corresponding to the small, intermediate, and large black hole phases. We find that the Page time depends sensitively on the spin parameter $a$, with smaller values of $a$ leading to an earlier occurrence of the Page time. Furthermore, for a fixed value of $a$, the Page time is observed to be earliest for the small black hole phase, followed by the intermediate black hole phase, and latest for the large black hole phase. Considering a black hole that is initially coupled to an external bath and subsequently isolated to undergo evaporation, we compute the Page curves in the small, intermediate, and large black hole regimes. We find that the presence of a first-order phase transition is imprinted on the corresponding Page curves through the appearance of a sharp discontinuity. This behavior is consistent with earlier results reported in \cite{Lin:2024gip}.

We then introduce a new ensemble by defining a thermodynamic quantity $\zeta$ conjugate to the spin parameter $a$ and analyze the corresponding thermodynamic behavior. We find that, in this ensemble, no first-order phase transition occurs, and the free energy exhibits only two distinct branches associated with the small and large black hole phases. We compute the Page curves for both phases and observe behavior qualitatively similar to that in the canonical ensemble, with the key difference that the Page time now depends on the parameter $\zeta$ rather than on $a$. We further investigate the Page curves in a setup where the black hole is initially coupled to an external bath and subsequently isolated to undergo evaporation. In this case, we find that the absence of a first-order phase transition implies that no discontinuity appears in the Page curve. We therefore conclude that a first-order phase transition can leave a clear imprint on the Page curve of a black hole, while its absence results in a smooth and continuous entropy evolution. 
\par For future directions, it will be interesting to extend this analysis to most general, charged and rotating black holes. The information paradox has been examined for flat 4D Kerr Newman black hole in \cite{yu2025islands}. In this paper we haven't considered the contribution from superradiance. However, it would be interesting to investigate the impact of superradiance on Page curve. Besides,it might be interesting to consider contributions from multiple island while calculating entanglement entropy.

\begin{acknowledgments}
D.D. and M.B.A. would like to thank Bidyut Hazarika for helpful discussions and assistance with certain calculations.
\end{acknowledgments}

%

\end{document}